\documentclass[10.5pt]{article}

\usepackage{mathptmx}
\usepackage{graphicx}
\usepackage{times}
\usepackage{amsmath, amsthm, amssymb}
\usepackage{url}
\usepackage{subfig}
\usepackage{hyperref}
\usepackage{algorithmic}
\usepackage{algorithm}
\usepackage{lscape}

\usepackage[latin1]{inputenc}
\usepackage[T1]{fontenc}
\usepackage{mathptmx}
\usepackage{url}

\usepackage{epsfig}
\usepackage{makeidx} 
\usepackage{graphicx}
\usepackage{epstopdf}

\begin{document}

\vspace{-0.8cm}
\begin{center}
\Large \textbf{\centerline{Economic inequality and mobility in kinetic models for social sciences}}

\vspace{0.5cm}
\normalsize

\normalsize{
Maria Letizia Bertotti$^1$, Giovanni Modanese$^2$
} \\
\vspace{5mm}
\textit{
$^1$ Faculty of Science and Technology, Free University of Bozen-Bolzano, Italy,
marialetizia.bertotti@unibz.it\\
$^2$ Faculty of Science and Technology, Free University of Bozen-Bolzano, Italy,
giovanni.modanese@unibz.it\\
}
\end{center}

\begin{abstract}
Statistical evaluations of the economic mobility of a society are more difficult than measurements of the income distribution, 
because they require to follow the evolution of the individuals' income for at least one or two generations. 
In micro-to-macro theoretical models of economic exchanges based on kinetic equations, 
the income distribution depends only on the asymptotic equilibrium solutions, 
while mobility estimates also involve the detailed structure of the transition probabilities of the model, 
and are thus an important tool for assessing its validity. Empirical data show a remarkably general negative correlation 
between economic inequality and mobility, whose explanation is still unclear. It is therefore particularly interesting to study this correlation 
in analytical models. In previous work we investigated the behavior of the Gini inequality index in kinetic models 
in dependence on several parameters which define the binary interactions and the taxation and redistribution processes: 
saving propensity, taxation rates gap, tax evasion rate, welfare means-testing etc. Here, we check the correlation of mobility with inequality
by analyzing the mobility dependence from the same parameters.
According to several numerical solutions, the correlation is confirmed to be negative.

\end{abstract}


\section{Introduction}
\label{Sec:Intro}
The rise of economic inequality has become, in recent times, cause for serious concern and the object of frequent debates. 
Inequality is increasing in several countries and also appears to be accompanied by a diminution of social mobility. This combination is especially harmful, 
because it frustrates the attempts of many individuals to move up the income ladder and improve their situation. 
Empirical surveys confirm the existence of a negative correlation between mobility and inequality \cite{Andrews D. Leigh. A., Corak}. 
It seems therefore very important to highlight the fundamental causes of this correlation.

Our aim in this paper is to explore these themes through mathematical methods, by providing a suitable definition of mobility 
and by comparing the possible scenarios compatible with different policies and conditions. In this connection, we introduce some indicators 
apt to measure various aspects of mobility within a family of kinetic-type models for the description of economic exchanges in a closed society.  
We investigate the relationship between economic inequality and mobility by analyzing income distributions emerging from different fiscal and welfare systems. 
The output of a large number of simulations shows clear evidence of a negative correlation between mobility and equality, 
in agreement with the qualitative features of real world situations.  The models also allow to study the simultaneous dependence of the Gini index 
and the mobility parameters on taxation rates and welfare means-testing, and to track variations in the mobility of different income classes.
In particular, it turns out (the simulations suggest) that there are, 
with reference to variables measuring welfare and taxation, 
lines on which both the Gini index and the mobility parameter are constant.

Our approach fits in with a complex system perspective, in the sense that
we find the emergence of collective macro-level patterns out of a whole of interactions occurring
at the micro level.
The need of considering, when dealing with socio-economic questions, 
a multiplicity of individual interactions
rather than a representative agent
is highlighted and advocated e.g. in \cite{Kirman A., Gallegati M., Arthur W.B.}.
For this purpose, agent based computational models have been typically 
developed and applied within the economics community, \cite{Tesfatsion L. Judd K.L.}.
On the other side, 
during the last two decades a number of physicists
have started to tackle economic and financial questions by using tools 
of statistical physics, see e.g. \cite{Dragulescu A. Yakovenko V.M., Sinha S. Chakrabarti B.K., 
Patriarca M. Chakraborti A.} and the recent review \cite{Chatterjee A.}.
The reason for that naturally arises from certain analogies existing between markets or populations of individuals 
who exchange money and particle systems or gases in which molecules collide exchanging energy.
In particular, as illustrated below, our methods and models allow, thanks to their characteristic features, 
the definition and investigation of a mobility concept. 

The remainder of the paper is organized as follows. 
Section $\ref{section:Framework}$ sets up a framework for the description of monetary exchanges in a closed society. 
In Section $\ref{section:An example of model family}$ a family of models is constructed within this framework. 
In the fourth section the concept of social mobility is shortly discussed and quantitative indicators to measure it are introduced. 
Hence, the correlation between the economic inequality and mobility is explored. 
Finally, a summary is contained in the last section.

\section{A discrete-type framework for the description of monetary exchanges in a closed society}
\label{section:Framework}
Moving from a stylized description of the complex of monetary exchanges which take place 
in a closed market society, we outline in this section 
a discrete-type framework suitable for its modelling.
This framework was proposed and discussed in our previous works \cite{Bertotti M.L., Bertotti M.L. Modanese G. 1, Bertotti M.L. Modanese G. 2},
but we give here a slightly different introduction.
The problem at issue 
concerns the formation of the income distribution curve 
and the economic inequality in a population of constant size.
If the variable $r \in [0,+\infty)$ denotes the individual income and $x(r)$ the density of the population having income $r$,
one could say that
the total population and the total income are given by
$\int_{0}^{+\infty} x(r) \, dr$ and $\int_{0}^{+\infty} r x(r) \, dr$ respectively.
However, it is quite natural and also convenient in view of
a numerical treatment of the problem
to divide the population into a finite number of classes. Toward this goal,
we consider the $n+1$ numbers $0 = r_0 < r_1< r_2 < ... < r_n$ with $r_n$ so large as to represent a sort of supremum of the
possible income of any individual in the given society. For simplicity we assume the density of the population
whose income belongs to the interval $[r_{i-1},r_i)$ to be constant and denote it by $x_i$ for $i = 1,...,n$. 
The fraction of individuals having income in $[r_{i-1},r_i)$ is then
\begin{equation}
X_i = \int_{r_{i-1}}^{r_i}  x_i \, dr = x_i (r_{i} - r_{i-1})
\label{relationPandx}
\end{equation}
and its income is
\begin{equation}
R_i = \int_{r_{i-1}}^{r_i} r x_i \, dr = x_i \, \frac{(r_{i}^2 - r_{i-1}^2)}{2}.
\label{relationRandx}
\end{equation}
We refer to the class of individuals with income in $[r_{i-1},r_i)$ as to the $i$-th class and to its individuals as to $i$-individuals.
An immediate consequence of $(\ref{relationPandx})$ and $(\ref{relationRandx})$ is that $R_i = X_i \, \tilde r_i$, where
$\tilde r_i = (r_{i-1}+r_{i})/2$ denotes the average income of individuals in the $i$-th class.

We assume that pairwise monetary exchanges
take place between single individuals. A "microscopic" description of this can be as follows: when an $h$-individual pays 
to a $k$-individual an amount of money $S$
(with $S < ( \tilde r_{i+1} -  \tilde r_{i})$ for all $i = 1, ..., n$), 
the $h$-th class  gets slightly poorer, while the $k$-th class gets slightly richer.
This induces a retrocession of a portion of the population of the $h$-th to the $(h-1)$-th class
and an advance of a portion of the $k$-th population to the $(k+1)$-th class.
Easy calculations show that, if the total population and the total income have to remain unchanged
as happens in the conservative case dealt with here,
then the variations of population in the $h$-th and in the $k$-th class are given by
\begin{equation}
\Delta X_h = - \frac{2S}{r_{h} - r_{h-2}} \qquad \hbox{and} \qquad \Delta X_k = - \frac{2S}{r_{k+1} - r_{k-1}}.
\label{deltaPhePk}
\end{equation}
How often monetary exchanges can occur can be postulated through the definition
of suitable terms $p_{h,k}$ for $h, k = 1, ... , n$, with
each $p_{h,k}$ 
accounting for the encounter frequency rate of individuals of the $h$-th and in the $k$-th class
and
expressing as well the probability
that  in an encounter between an $h$-individual and a $k$-individual,
the one who pays is the $h$-individual. 
The possibility that none of the two pays has to be taken into account as well.
Hence, for any $h$ and $k$ in $\{1, ... , n\}$, the inequalities $0 \le p_{h,k} \le 1$ and $p_{h,k} + p_{k,h} \le 1$ have to hold true.
Apart from that, the values of the $p_{h,k}$ should be determined
based on phenomenological observations.
A specific choice for the values of the $p_{h,k}$
will be done in the third section of this note.

Beside direct money exchanges, we
also assume the existence of a process of taxation and redistribution.
If we imagine that the money received form the $k$-individual is subject to taxation with a tax rate equal to $\tau_k \in [0,1)$,
supposing (at least as a first case study) that the $k$-individual pays the due tax,
we observe that the entire operation is equivalent to one in which
the $h$-individual pays a quantity $S$,
the $k$-individual earns a quantity $S \, (1 - \tau_k)$,
and the quantity $S \, \tau_k$ goes to the government.
We then imagine, here again postulating 
an irreproachable situation, that the government redistributes
the tax revenue (which is just $S \, \tau_k$ for the single mentioned interaction) to the entire population through the provision of welfare services:
education, health assistance and so on.
At this step, we observe that in fact the government intervention can be ignored, since the mechanism can be
equivalently described as follows: in correspondence to
the mentioned {\it direct} interaction between an $h$-individual and a $k$-individual, individuals of all classes
receive an amount of money corresponding to a returned tax portion.
In other words, the taxation and redistribution effect is equivalent to the effect of an {\it indirect} interaction 
between the paying $h$-individual and
all others.\footnote{\ We point out that actually, for a technical reason, in the following we will 
assume that no redistribution goes to individuals of the $n$-th class.
Indeed, if an individual of the $n$--th class would receive some money, the possibility would arise 
for him to advance to a higher class, which is impossible.}
Of course, the indirect interactions too contribute to the possibility of class change of individuals.

In order to formalize the overall mechanism described so far we introduce

{$-$}  the {\slshape direct transition probabilities} 
$C_{hk}^i \in [0,1]$ for $i, h, k \in \{1, ..., n\}$,
satisfying
\begin{equation}
\sum_{i=1}^n C_{hk}^i = 1 \qquad \hbox{for any} \  h, k \in \{1, ..., n\},
\label{A probability}
\end{equation}
where $C_{hk}^i$ 
expresses the probability 
that an individual of the $h$-th class 
will belong to the 
$i$-th class after a direct interaction with an
individual of the $k$-th class,

and

{$-$}  the {\slshape indirect transition variations} 
$T_{[hk]}^i  \, :   {\bf R}^n \to {\bf R}$ for $i, h, k \in \{1, ..., n\}$,
where each $T_{[hk]}^i(X)$ with $X=(X_1,...,X_n) \in {\bf R}^n$ is a continuous function, satisfying
\begin{equation}
\sum_{i=1}^n T_{[hk]}^i(X) = 0 \qquad \hbox{for any} \  h, k \in \{1, ..., n\}
\ \hbox{and} \ X \in {\bf R}^n,
\label{T probability}
\end{equation}
accounting for the indirect interactions and expresses the variation in the $i$-th class
due to an interaction between an $h$-individual and a $k$-individual.

As will be shown in the following section,
the construction of both the parameters $C_{hk}^i$ and the functions $T_{[hk]}^i$ 
is accomplished combining
the variation terms $(\ref{deltaPhePk})$ and the probabilities $p_{h,k}$. 

The variation in time of the number of individuals belonging to the $n$ income classes
is described by the following balance differential equations:
\begin{equation}
{{d X_i} \over {d t}} =  
\sum_{h=1}^n \sum_{k=1}^n C_{hk}^i X_h X_k 
- \sum_{h=1}^n \sum_{k=1}^n C_{ik}^h X_i X_k  
+ \sum_{h=1}^n \sum_{k=1}^n T_{[hk]}^i(X) X_h X_k
\qquad \hbox{for} \  i \in \{1, ..., n\},
\label{evolution eq of P}
\end{equation}
which, in view of $(\ref{A probability})$, can be simplified to take the form:
\begin{equation}
{{d X_i} \over {d t}} =  
\sum_{h=1}^n \sum_{k=1}^n {\Big (} C_{hk}^i + T_{[hk]}^i(X) {\Big )} \, X_h X_k
- X_i \sum_{k=1}^n X_k
\qquad \hbox{for} \  i \in \{1, ..., n\}.
\label{evolution eq}
\end{equation}
They are nonlinear ordinary differential equations.

\section{An example of model family}
\label{section:An example of model family}
In order to construct a specific model or model family, we start by fixing specific values for the $p_{h,k}$.
We take them as 
$$
p_{h,k} = \min \{\tilde r_{h},\tilde r_{k}\}/{4 \tilde r_{n}},
$$
with the exception of the terms
$p_{j,j} = {\tilde r_{j}}/{2 \tilde r_{n}}$ for $j = 2, ..., n-1$,
$p_{h,1} = {\tilde r_1}/{2 \tilde r_{n}}$ for $h = 2, ..., n$, 
$p_{n,k} = {\tilde r_{k}}/{2 \tilde r_{n}}$ for $k = 1, ..., n-1$,
$p_{1,k} = 0$ for $k = 1, ..., n$
and
$p_{hn} = 0$ for $h = 1, ..., n$.
This choice is suggested by the consideration that poorer individuals typically
earn and pay smaller amounts than richer ones and do so less frequently.
It guarantees a degree of heterogeneity among
people of different classes, also in connection with
their interactions with others.

Recalling now the formulae $(\ref{deltaPhePk})$,
we assume that
the only possibly nonzero elements among the $C_{hk}^i$ are:
\begin{eqnarray}
C_{i+1,k}^{i} & = 
                  & p_{i+1,k} \, \frac{2S \, (1-\tau_k) }{r_{i+1} - r_{i-1}}, \nonumber \\
C_{i,k}^i & = 
            & 1 - \, p_{k,i} \, \frac{2S \, (1-\tau_i)}{r_{i+1} - r_{i-1}} 
               - \, p_{i,k} \, \frac{2S \, (1-\tau_k)}{r_{i} - r_{i-2}}, \nonumber \\
C_{i-1,k}^i & = 
               & p_{k,i-1} \, \frac{2S \, (1-\tau_{i-1})}{r_{i} -  r_{i-2}}. 
\label{choiceforCnewapproach}
\end{eqnarray} 
We stress that the expression for $C_{i+1,k}^{i}$ in $(\ref{choiceforCnewapproach})$ holds true for $i \le n-1$ and $k\le n-1$;
the second addendum of the expression for $C_{i,k}^i$ is effectively present only 
provided $i \le n-1$ and $k \ge 2$, while its third addendum is present
only provided $i \ge 2$ and $k \le n-1$; the expression for $C_{i-1,k}^i$ holds true for $i \ge 2$ and $k\ge 2$.

\smallskip

As for the $T_{[hk]}^i$, we first introduce certain coefficients $w_i$ for $i = 1, ..., n$, which denote the weights
of a differently distributed welfare.
In the following we will take the $r_j$ (and consequently the $\tilde r_j$) linear in $j$: $r_j = c j$ for some constant $c>0$.
Then, a conceivable expression for the weights $w_i$, according to which,  
the $w_j$ decrease linearly as functions of $j$, 
is given by
\begin{equation}
w_j = \tilde r_{n+1-j} + \frac{2}{n-1} \, \gamma \, \bigg(j - \frac{n+1}{2}\bigg) \, (\tilde r_{n} - \tilde r_{1}),
\label{formula welfare}
\end{equation}
with $\gamma \in (0,1/2)$. 
The smaller $\gamma\in (0,1/2)$, the larger is the difference
$w_1 - w_n$. Indeed, $w_1 - w_n = (\tilde r_{n} - \tilde r_{1}) \, (1 - 2 \gamma)$.
The value $\gamma = 1/2$ also is admitted, but if $\gamma = 1/2$, then
$w_j$ has the same value for each $j = 1, ..., n$.

We can now express the $T_{[hk]}^i$ as 
\begin{equation}
T_{[hk]}^i(X) =  
U_{[hk]}^i(X) + V_{[hk]}^i(X),
\label{T+U}
\end{equation}
where\footnote{\ 
In $(\ref{U_{[hk]}^i(P)})$,
$h >1$ 
and
the terms 
involving the index $i-1$ [respectively, $i+1$]
are effectively present only provided $i-1 \ge 1$ 
[respectively, $i+1 \le n$]. In other words:
the $1^o$ term into parentheses on the r.h.s. of (\ref{U_{[hk]}^i(P)}) is present for $2 \le i \le n$;
the $2^o$ term into parentheses on the r.h.s. of (\ref{U_{[hk]}^i(P)}) is present for $1 \le i \le n-1$.
}
\begin{equation}
U_{[hk]}^i(X) =  
\frac{p_{h,k} \, 2S \, \tau_k}{\sum_{j=1}^{n} w_j X_{j}} {\bigg (}  \frac{w_{i-1} X_{i-1}}{r_i - r_{i-2}} -  \frac{w_i X_{i}}{r_{i+1} - r_{i-1}} {\bigg )}
\label{U_{[hk]}^i(P)}
\end{equation}
represents
the variation corresponding to the advancement
from a class to the subsequent one, due to the benefit of taxation
and\footnote{\ 
In $(\ref{V_{[hk]}^i(P)})$,
$h >1$ 
and
the terms 
involving the index $i-1$ [respectively, $i+1$]
are effectively present only provided $i-1 \ge 1$ 
[respectively, $i+1 \le n$]. In other words:
the $1^o$ term into parentheses on the r.h.s. of (\ref{V_{[hk]}^i(P)}) is present for $1 \le i \le n-1$;
the $2^o$ term into parentheses on the r.h.s. of (\ref{V_{[hk]}^i(P)}) is present for $2 \le i \le n$.
}
\begin{equation}
V_{[hk]}^i(X)
=  p_{h,k} \, 2S \, \tau_k \, \frac{\sum_{j=1}^{n-1} w_j X_{j}}{\sum_{j=1}^{n} w_j X_{j}} \, 
{\bigg (} 
\frac{\delta_{h,i+1}}{r_h - r_{i-1}} \, - \, \frac{\delta_{h,i}}{r_h - r_{i-2}}
{\bigg )},
\label{V_{[hk]}^i(P)}
\end{equation}
with $\delta_{h,k}$ denoting the {\slshape Kronecker delta}, 
represents the variation corresponding to the retrocession
from a class to the preceding one, due to the payment of some tax.

We notice here that the effective amount of money representing taxes, which is paid 
in correspondence to a payment of $S (1 - \tau_k)$
and is then redistributed 
is $S \, \tau_k \,({\sum_{j=1}^{n-1} w_j X_{j}})/({\sum_{j=1}^{n} w_j X_{j}})$
instead of $S \, \tau_k$. This is in agreement with the fact that, due to the bound on the admissible income, individuals of the $n$-th class
do not benefit from the redistribution (see Note $1$).

Finally, to fix ideas, we take $S = 1$, 
\begin{equation}
r_j = 25 \, j,
\label{average incomes}
\end{equation}
and
\begin{equation}
\tau_j = \tau_{min} +  \frac{j - 1}{n-1} \, (\tau_{max} - \tau_{min}),
\label{progressivetaxrates}
\end{equation}
for $j = 1, ... , n$, where $\tau_{min}$ and $\tau_{max}$ respectively denote the minimum and maximum tax rate.
Still, the value of $\gamma$, $\tau_{min}$ and $\tau_{max}$
have to be fixed. 
Hence, the equations $(\ref{evolution eq})$ describe a family of models rather than a single model.

Finding analytical solutions for these equations is in practice hopeless. But,
as will be shown in the next section, several relevant facts can be understood through simulations. 
Before passing to numerical results, we would like to mention the following two properties
of $(\ref{evolution eq})$. For both of them the proof of the corresponding result established
in \cite{Bertotti M.L.} and referring in fact to a slightly different model
can be easily adapted.

\noindent {\bf {Well posedness of the Cauchy problem}}.
For fixed values of the parameters $\gamma$, $\tau_{min}$ and $\tau_{max}$,
in correspondence to any initial condition $X_0 = (X_{01} , \ldots , X_{0n})$, 
for which $X_{0i} \ge 0$ for all $i = 1, ... , n$ and $\sum_{i=1}^n X_{0i} = 1$,
a unique solution $X(t) = (X_1(t),\ldots,X_n(t))$ of $(\ref{evolution eq})$ exists,
which is defined for all $t \in [0,+\infty)$, satisfies $X(0) = X_0$ and also
\begin{equation}
X_{i}(t) \ge 0 \quad \hbox{for} \ i = 1, ... , n \qquad \hbox{and} \qquad \sum_{i=1}^n X_{i}(t) = 1 \quad  \hbox{for all} \ t \ge 0 \, . 
\label{solution in the future}
\end{equation}

\noindent {\bf {Conservation of the total income $\mu$}}.
For fixed values of the parameters $\gamma$, $\tau_{min}$ and $\tau_{max}$,
the scalar function
$\mu(X)=\sum_{i=1}^n \tilde r_i X_i$, expressing the total (and mean) income, 
is a first integral for the system $(\ref{evolution eq})$. 

\medskip

From now on, we report on the outputs of a large number of simulations. To run them, we fixed $n = 15$.
The first fact one observes is the following one.

\noindent {\bf {Uniqueness, for any fixed value of $\mu$, of the asymptotic stationary distribution}}.
For fixed values of the parameters $\gamma$, $\tau_{min}$ and $\tau_{max}$,
for any fixed value $\mu \in [\tilde r_1,\tilde r_n]$, an equilibrium of $(\ref{evolution eq})$ exists, to which
all solutions of $(\ref{evolution eq})$, whose initial conditions $X_0 = (X_{01} , \ldots , X_{0n})$ satisfy
$X_{0i} \ge 0$ for all $i = 1, ... , n$, $\sum_{i=1}^n X_{0i} = 1$, and $\sum_{i=1}^n \tilde r_i X_{0i} = \mu$
tend asymptotically as $t \to +\infty$.
In other words, a one-parameter family of asymptotic stationary distributions exists,
the parameter being the total income value. 

Other properties 
can be established from comparisons between different models of the family under consideration.
In previous work (\cite{{Bertotti M.L. Modanese G. 4, Bertotti M.L. Modanese G. 5}}) we investigated 
the effects on the shape of the asymptotic income distribution curve and the economic inequality
produced by the adoption of fiscal systems with different spread between the maximum and the minimum tax rates
and also 
produced by the introduction of differently weighted welfare measures. Briefly, the simulations suggested the following properties.

\noindent {\bf {Dependence of the asymptotic stationary distribution on the tax rate difference $\tau_{max} - \tau_{min}$}}.
The profile of the asymptotic stationary distribution depends on the difference between the maximum and the minimum tax rate:
if $\tau_{max} - \tau_{min}$ is enlarged, while all other data are kept unchanged, then an increase of the
fraction of individuals belonging to the middle classes (to the detriment of those in
the poorest and richest classes) can be detected at the asymptotic equilibrium.

\noindent {\bf {Dependence of the asymptotic stationary distribution on differently weighted welfare measures}}. 
The profile of the asymptotic stationary distribution depends on the parameter $\gamma$ characterising the
welfare measures. When $\gamma$ decreases and all other data are kept unchanged, at the
asymptotic equilibrium an increase of the fraction of individuals belonging to the middle classes
can be detected (and, correspondingly, a decrease of those in the poorest and richest classes).

In this connection we evaluated for the models under consideration
the Gini index $G$ and the tax revenue $T \!R$,
whose definition is recalled next.
Let the Lorenz curve represent (on the $y$ axis) the cumulative percentage of the total income of a population  
earned by the bottom percentage of individuals (on the $x$ axis).
The Gini index $G$ corresponds to the ratio $A_1/A_2$ of the area $A_1$ between the line of perfect equality $y = x$ and the Lorenz curve 
and the total area $A_2$ under the line of perfect equality. It takes values in $[0,1]$, $0$ representing complete equality and $1$ maximal inequality. 
The tax revenue is the total amount of tax collected in the unit time and redistributed as welfare provisions:
\begin{equation}
T\!R = \sum_{h=1}^{n} \, \sum_{k=1}^{n} \,\sum_{j=1}^{n - 1} \, \, p_{hk} \, S \tau_k \, 
\frac{w_j {{\tilde X}_j}}{(\sum_{i=1}^{n} w_i {{\tilde X}_i})} \, {{\tilde X}_h}{{\tilde X}_k},
\end{equation}
with ${\tilde X}_i$ denoting the fraction of individuals in the $i$-th class at equilibrium. 

\section{Social mobility and its correlation with economic inequality}
\label{section:Mobility $...$}

This paper focuses upon an important socio-economic issue: social mobility.\footnote{\ The expressions social mobility and economic mobility 
are used here to designate the same concept.}

\subsection{Introductive remarks}

Social mobility gives a measure of the chances individuals have to improve their economic status through education and work. 
It is known that a high level of inequality in a society is especially harmful when it is complemented by a diminution of social mobility,
see e.g. \cite{Andrews D. Leigh. A., Corak}.
Social mobility has been historically an important component of the ``American Dream'' (and of its emulations), 
thus making the high inequality level present in the US more tolerable. 
Today, several observers point out a diminution of social mobility, as seen for instance from the growing difficulties that 
the sons and daughters of the poor and middle classes have to access the best universities and the best paid jobs.
As mentioned in the introduction, quantitative empirical surveys show that mobility is negatively
correlated with inequality, i.e. 
more unequal societies also tend to have less mobility.

Studying the correlation between economic inequality and mobility is of particular interest
because
statistical evaluations of the economic mobility of a society are more difficult than measurements of the income distribution.
This is due to the fact that
they require to follow the evolution of the individuals' income for at least one or two generations. 
Correspondingly, in micro-to-macro theoretical models of economic exchanges based on kinetic equations, 
the income distribution depends only on the asymptotic equilibrium solution of the equations, while
mobility estimates also involve the detailed structure of the transition probabilities of the model.

In our models there are no generations and it is impossible to compare the income of one individual with that of his or her parents. 
Individuals are supposed to live and work indefinitely, or at least for the time it takes for the system to reach its dynamical equilibrium. 
During this time the income distribution evolves from some initial conditions to 
a stationary state, in correspondence to which
the number of individuals belonging to a certain class 
remains constant in time: at equilibrium, the total rate of individuals leaving that class is equal in absolute value to the total rate of individuals arriving from other classes. 
This means that in the equations $(\ref{evolution eq})$,
whose right hand side can now be written in a simplified form in view of $(\ref{solution in the future})$,
one has $\dot X_i=0$.
Equivalently, the equilibrium populations are solutions of the nonlinear equations 
$$
\sum_{h=1}^n \sum_{k=1}^n {\Big (} C_{hk}^i + T_{[hk]}^i(X) {\Big )} \, X_h X_k 
- X_i = 0
\qquad \hbox{for} \  i \in \{1, ..., n\},
$$
with $C_{hk}^i$ and $T_{[hk]}^i(X)$
as in $(\ref{choiceforCnewapproach})$ and $(\ref{T+U})$. 

The equilibrium populations can be obtained numerically as asymptotic solutions of the full evolution equations. 
We have evidence from several numerical solutions that the approach to equilibrium is exponential in time, 
in the sense that the ``Cauchy convergence norm''
converges exponentially to zero as $t \to \infty$ (see \cite{Bertotti M.L. Modanese G. 4}).

Below we introduce quantitative indicators suitable to define the mobility in our models. Then, we check the 
correlation of mobility with inequality. To this end, we analyze the dependence of mobility on parameters such as
the taxation rates gap and welfare means-testing; we find interesting relations between the mobility and the inequality indicators.

\subsection{Mobility in our model: detailed calculation and dependence on the model parameters}

The mobility evaluated in this paper is always meant at equilibrium, when the rates of the individuals entering and leaving an income class are the same 
(but of course differ from class to class). We do not consider long-term trends, characterized by an increase or diminution of the total income of the society. 
In our models the social mobility is essentially a short-term mobility, which gives each individual 
a probability to go up or down the income ladder. 
A large value of the mobility should be regarded as positive for the economy, 
because it acts as a stimulus and as a reward for the individuals who are more capable and hard-working. 
While in the average the upward mobility and downward mobility are the same at equilibrium, we can imagine that those who go up are economically stronger 
than those who go down. This might be evidenced, in an improved version of the model, by introducing some heterogeneity factors, and so labelling 
the individuals not only with their income class $i$ but possibly also with some ``fitness'' index related to education, number of economic links, etc.

To start with, we define
the probabilities for an individual belonging to the income class $i$, for $i = 2, ..., n-1$, to be promoted to the upper class 
following a direct interaction or an indirect interaction (i.e., through welfare provisions) 
respectively 
as
\begin{equation}
{P_{i,exchanges (individual)}} = \frac{S}{{{r_{i + 1}} - {r_i}}} \,\sum_{k=1}^{n} \, {{p_{k,i}}(1-{\tau _i}){{\tilde X}_k}}
\label{Piindividualexc}
\end{equation}
and
\begin{equation}
{P_{i,welfare (individual)}} = \frac{S}{{{r_{i + 1}} - {r_i}}} 
\, \frac{w_i}{\sum_{j=1}^n w_j {\tilde X}_{j}} \sum_{h=1}^{n} \, \sum_{k=1}^{n} \, {{p_{h,k}}{\tau _k}{{\tilde X}_h}{{\tilde X}_k}}.
\label{Piindividualwel}
\end{equation}
We restrict our attention to all classes but the first and the last one, in order to avoid possible boundary effects. 
By summing the probabilities
$(\ref{Piindividualexc})$ and $(\ref{Piindividualwel})$
we obtain the total probability $P_{i (individual)}$ for a single individual advancement
$$
{P_{i (individual)}} = {P_{i,exchanges (individual)}}  + {P_{i,welfare (individual)}},
$$
which is the base for the calculations that follow.

We then define ``averaged'' class probabilities of being promoted to the upper class
(the one-step transition probabilities for the class $i$)
respectively due to direct or indirect interactions as 
\begin{equation}
{P_{i,exchanges (class)}} = \frac{1}{(1 - {\tilde X}_{1} - {\tilde X}_{n})} \, \frac{S}{{{r_{i + 1}} - {r_i}}} \,\sum_{k=1}^{n} \, {{p_{k,i}}(1-{\tau _i}){{\tilde X}_k}{{\tilde X}_i}} 
\label{Piclassexc}
\end{equation}
and
\begin{equation}
{P_{i,welfare (class)}} = \frac{1}{(1 - {\tilde X}_{1} - {\tilde X}_{n})} \, \frac{S}{{{r_{i + 1}} - {r_i}}} 
\, \frac{{w_i {\tilde X}_{i}}}{\sum_{j=1}^n w_j {\tilde X}_{j}} \sum_{h=1}^{n} \, \sum_{k=1}^{n} \, {{p_{h,k}}{\tau _k}{{\tilde X}_h}{{\tilde X}_k}}
\label{Piclasswel}
\end{equation}
for $i = 2, ..., n-1$.

We have first introduced similar quantities
in our paper \cite{Bertotti M.L. Modanese G. 4}, where we computed the ratio between 
the term due to binary exchanges (the quadratic part of the evolution equation) and the term due to tax redistribution (the cubic part). 
This ratio gives a measure of the ``liberalism'' of the society. We found that with equally distributed welfare the ratio stays, quite surprisingly, 
almost constant when we pass from the middle classes to the rich classes, although one might expect welfare provisions to be irrelevant 
for the class advancement of the super-rich. Let us now call
$$
{P_{i (class)}} = {P_{i,exchanges (class)}}  + {P_{i,welfare (class)}}.
$$
the sum of the probabilities $(\ref{Piclassexc})$ and $(\ref{Piclasswel})$.

\begin{figure}[ht]
\begin{center}
\includegraphics[width=6.5cm,height=3.0cm]{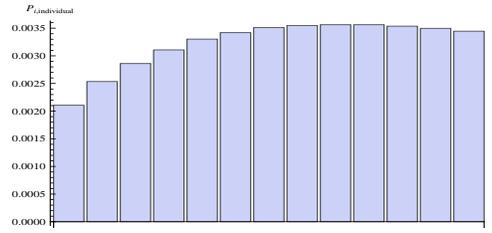}
\caption{\small{Individual probability of single-class advancement $P_{i(individual)}$ obtained from a numerical simulation with 15 income classes. 
The probability has been computed from the equilibrium population and from the coefficients $p_{i,k}$, $\tau_i$, according to Eq. $(\ref{evolution eq})$, 
in a case where $\tau_{min}=30\%$, $\tau_{max}=45\%$ and the resulting Gini index is $G=0.368$. Note that the rich classes have the largest individual mobility. 
The initial conditions for the numerical simulation are such that 
most of the wealth initially belongs to the low and middle classes.}}
\label{P-indiv-368}
\end{center}
\end{figure}

\begin{figure}[ht]
\begin{center}
\includegraphics[width=6.5cm,height=3.0cm]{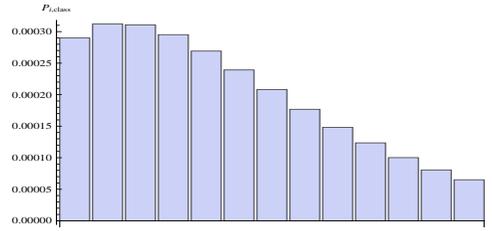}
\caption{\small{Histogram of the class mobility $P_{i(class)}$ with the same parameters as in Fig.\ 1. This is obtained from the histogram of the individual mobility, 
weighing each column with the normalized population of the corresponding class. Since the low and middle classes are much more populated than the rich classes 
(the first 7 or 8 classes typically comprise 4/5 of the entire population), the weights affect strongly the relative contributions of the individual mobilities of Fig.\ 1. 
The total mobility $M$ is determined by a weighted average.}}
\label{P-class-368}
\end{center}
\end{figure}

If we make an histogram of $P_{i (individual)}$ in a typical situation with Gini index $G=0.368$,  
we see that it increases with income and attains its largest values in correspondence of the richest classes (see Fig. $\ref{P-indiv-368}$). 
We know, however, that 
in real world income distributions, the fifth quintile generally includes a consistent share of the classes, 
since the richest classes are scarcely populated
and most of the population concentrates in the low- and middle-income classes. 
This also happens in our models provided the
initial conditions are such that
most of the wealth initially belongs to the low and middle classes.
For instance, in the example with the parameters of Fig. $\ref{P-indiv-368}$, 
the histogram of mobility is shown in Fig. $\ref{P-class-368}$. So, while some aspects of the mobility of the rich are interesting in principle, 
because they are related to the formation of the Pareto fat tail in the distribution,
the value of the mobility of the rich classes does not have a strong influence on the average mobility of all classes. 

Our main measure of mobility expresses
the collective probability of class advancement of all classes from the $2$-th to the $(n - 1)$-th one. Indeed, we consider
\begin{equation}
{P_{exchanges (collective)}} = \frac{1}{(1 - {\tilde X}_{1} - {\tilde X}_{n})} \, \frac{S}{{{r_{i + 1}} - {r_i}}} \, \sum_{i=2}^{n-1} \,\sum_{k=1}^{n} \, {{p_{k,i}}(1-{\tau _i}){{\tilde X}_k}{{\tilde X}_i}} 
\end{equation}
and
\begin{equation}
{P_{welfare (collective)}} = \frac{1}{(1 - {\tilde X}_{1} - {\tilde X}_{n})} \, \frac{S}{{{r_{i + 1}} - {r_i}}} \, \sum_{i=2}^{n-1}
\, \frac{{w_i {\tilde X}_{i}}}{\sum_{j=1}^n w_j {\tilde X}_{j}} \sum_{h=1}^{n} \, \sum_{k=1}^{n} \, {{p_{h,k}}{\tau _k}{{\tilde X}_h}{{\tilde X}_k}}
\end{equation}
and take the sum 
\begin{equation}
M 
= {P_{exchanges (collective)}}  + {P_{welfare (collective)}}
\label{mobility}
\end{equation}
as the parameter which represents the mobility.

%
%

\begin{figure}[t]
\begin{center}
\includegraphics[width=6.5cm,height=1.75cm]{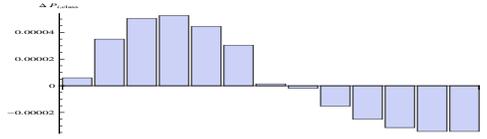}
\caption{\small{Variation $\Delta P_{i(class)}$ of the class mobility when the tax rates change from $30\% - 45\%$ to $30\% - 60\%$, 
and the Gini index correspondingly decreases from $G=0.368$ to $G=0.338$.}}
\label{DeltaP-class}
\end{center}
\end{figure}

\begin{figure}[t]
\begin{center}
\includegraphics[width=6.5cm,height=1.75cm]{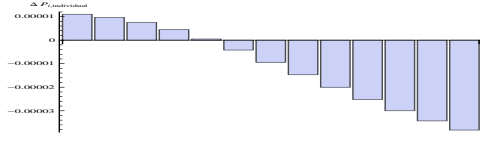}
\caption{\small{Variation $\Delta P_{i(individual)}$ of the individual mobility when the tax rates change from $30\% - 45\%$ to $30\% - 60\%$, 
and the Gini index correspondingly decreases from $G=0.368$ to $G=0.338$.}}
\label{DeltaP-indiv}
\end{center}
\end{figure}

Now, we change slightly one of the parameters of the model, 
in order to increase equality and decrease the Gini index. For instance, we pass from taxation rates $30\%-45\%$ to $30\%-60\%$, 
leaving unchanged the parameters defining the direct exchanges and the welfare redistribution parameter (here we have $\gamma=0.5$, 
which means uniform welfare distribution). 
This leads to a Gini coefficient $G=0.338$, to a variation $\Delta P_{i (class)}$ 
as in Fig. $\ref{DeltaP-class}$, 
and to an average variation $\Delta M = 5.7 \cdot 10^{-5}$. 
It is clear from the histogram that the mobility variation for the middle classes is positive, and that it prevails in the average. The histogram of the individual mobility variations 
shows an even sharper diminution of the mobility of the rich classes, see Fig. $\ref{DeltaP-indiv}$. 
In fact, one could verify that also the percentage mobility variation is always larger for the rich classes, 
both for ${P_{(individual)}}$ and ${P_{(class)}}$. 
A similar behavior is found when the negative variation of the Gini index is obtained 
with a change of the welfare redistribution parameter, as to make it more favorable to the low and middle classes. Several trials of this kind 
confirm the positive correlation between equality and mobility. 
In the next section  
we shall consider variations of the Gini index and mobility in correspondence to contemporary variations in tax rates and welfare distribution parameters, 
which again confirm this positive correlation.

%

\subsection{Gini index $G$ and mobility $M$ in dependence on two variables: level lines and correlation}

\begin{figure}[h]
\begin{center}
\includegraphics[width=7.0cm,height=5.125cm]{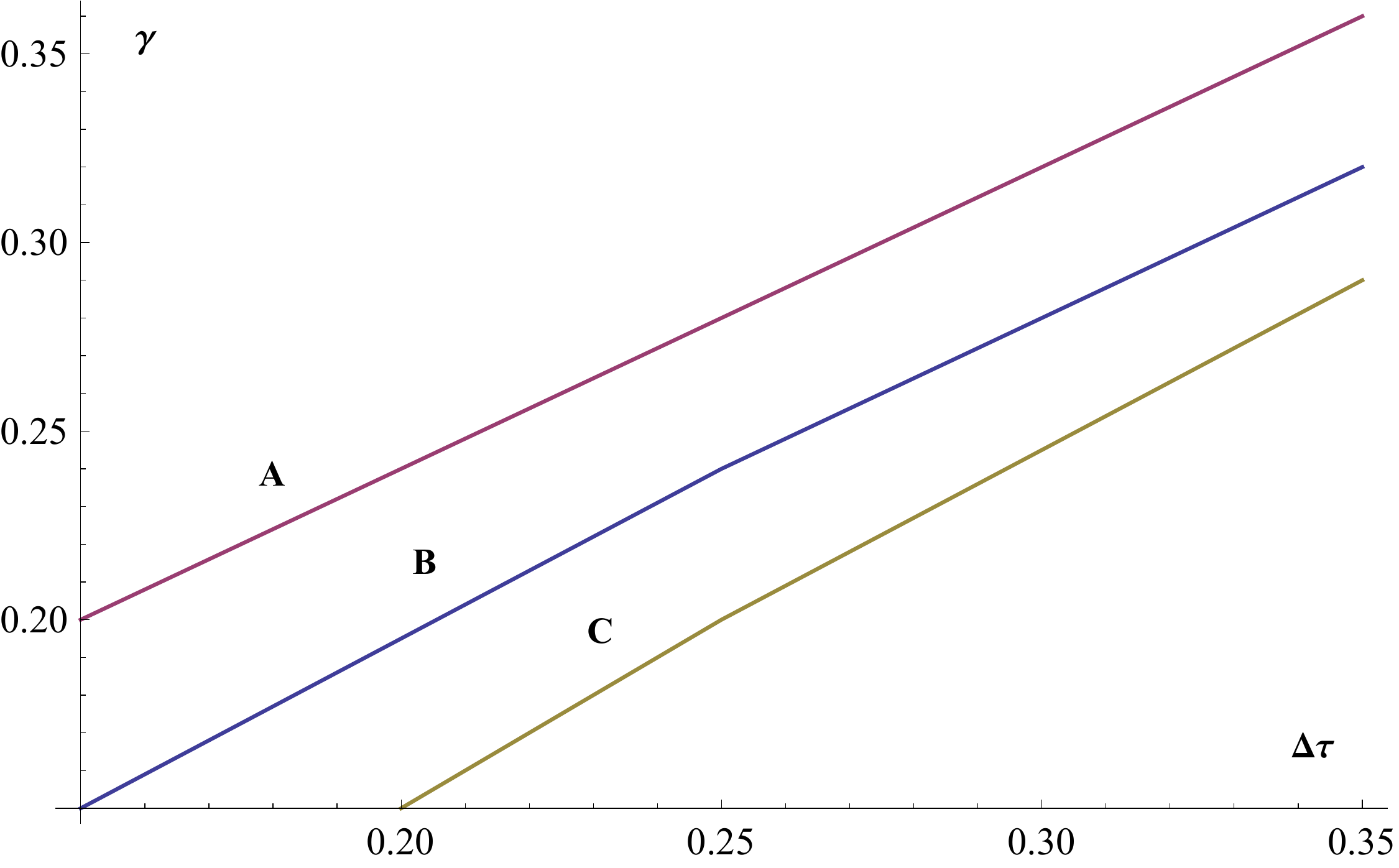}
\caption{Level lines in the plane $\Delta \tau - \gamma$ corresponding to the following values of Gini index and mobility. 
{\bf A}: $G = 0.341$, $M = 0.00270$. {\bf B}: $G = 0.338$, $M = 0.00271$. {\bf C}: $G = 0.335$, $M = 0.00272$. 
Numerical data for the three lines are given resp.\ in Table \ref{t1}, \ref{t2}, \ref{t3}.}
\label{Levellines}
\end{center}
\end{figure}

If we compute the values of $G$ in dependence on the two variables $\Delta \tau = \tau_{max} - \tau_{min}$ and $\gamma$, 
we can identify on a cartesian plane $\Delta \tau - \gamma$ a family of ``level lines of $G$'' corresponding to constant values of $G$. 
Similarly, we can find in the same plane the level lines of $M$.
In principle, one would expect the two sets of level lines to be distinct. As a matter of fact, it is impossible 
to write an explicit functional dependence of $G$ and $M$ on $\Delta \tau$ and $\gamma$, and
the definitions of the two quantities are completely different. In particular, $G$ depends only on the $\tilde X_i$,
while $M$ also depends on the encounter frequency rates $p_{hk}$, which are largely arbitrary.
What is remarkable, however, is that the level lines of $G$ and $M$ coincide up to ca. $1$ part in $10^3$, 
at least in the region of the plane $\Delta \tau - \gamma$ 
which we have explored through the simulations. The increments of $G$ and $M$ are always opposite, so the negative correlation 
between them is strongly confirmed also in simultaneous dependence on these two variables.
Fig. $\ref{Levellines}$ represents, for instance, the three coincident level lines of $G$ and $M$ which start from the points 
$(\Delta \tau,\gamma)$ equal to $(0.15,0.20)$, $(0.15,0.15)$ and $(0.20,0.15)$. 
The corresponding values of $\Delta \tau$, $\gamma$, $G$ and $M$ are reported in Tab. $\ref{t1} - \ref{t3}$.

\begin{table}[!h]
\centering
\begin{tabular}{ccccc}
\hline\noalign{\smallskip}
${\bf \tau_{min} - \tau_{max}}$ & \ ${\boldsymbol \Delta \tau}$ & \ ${\bf \gamma}$ & \ ${\bf G}$ & \ ${\bf M}$ \\
\noalign{\smallskip}\hline\noalign{\smallskip}
30 - 45 & \ 0.15 & \ 0.20 & \ 0.341 & \ 0.002700 \\
25 - 50 & \ 0.25 & \ 0.28 & \ 0.341 & \ 0.002704 \\
20 - 55 & \ 0.35 & \ 0.36 & \ 0.341 & \ 0.002707 \\
\noalign{\smallskip}\hline
\end{tabular}
\caption{Data for level line A.}
\label{t1}
\end{table}
\begin{table}[!h]
\centering
\begin{tabular}{ccccc}
\hline\noalign{\smallskip}
${\bf \tau_{min} - \tau_{max}}$ & \ ${\boldsymbol \Delta \tau}$ & \ ${\bf \gamma}$ & \ ${\bf G}$ & \ ${\bf M}$ \\
\noalign{\smallskip}\hline\noalign{\smallskip}
30 - 45 & \ 0.15 & \ 0.15 & \ 0.338 & \ 0.002712 \\
25 - 50 & \ 0.25 & \ 0.24 & \ 0.338 & \ 0.002714 \\
20 - 55 & \ 0.35 & \ 0.32 & \ 0.338 & \ 0.002717 \\
\noalign{\smallskip}\hline
\end{tabular}
\caption{Data for level line B.}
\label{t2}
\end{table}
\begin{table}[!h]
\centering
\begin{tabular}{ccccc}
\hline\noalign{\smallskip}
${\bf \tau_{min} - \tau_{max}}$ & \ ${\boldsymbol \Delta \tau}$ & \ ${\bf \gamma}$ & \ ${\bf G}$ & \ ${\bf M}$ \\
\noalign{\smallskip}\hline\noalign{\smallskip}
27.5 - 47.5 & \ 0.20 & \ 0.15 & \ 0.335 & \ 0.002723 \\
25 - 50 & \ 0.25 & \ 0.20 & \ 0.335 & \ 0.002723 \\
20 - 55 & \ 0.35 & \ 0.28 & \ 0.335 & \ 0.002727 \\
\noalign{\smallskip}\hline
\end{tabular}
\caption{Data for level line C.}
\label{t3}
\end{table}

\subsection{Mobility in the kinetic model of a physical gas}

\begin{figure}[h]
\begin{center}
\includegraphics[width=7.0cm,height=5.125cm]{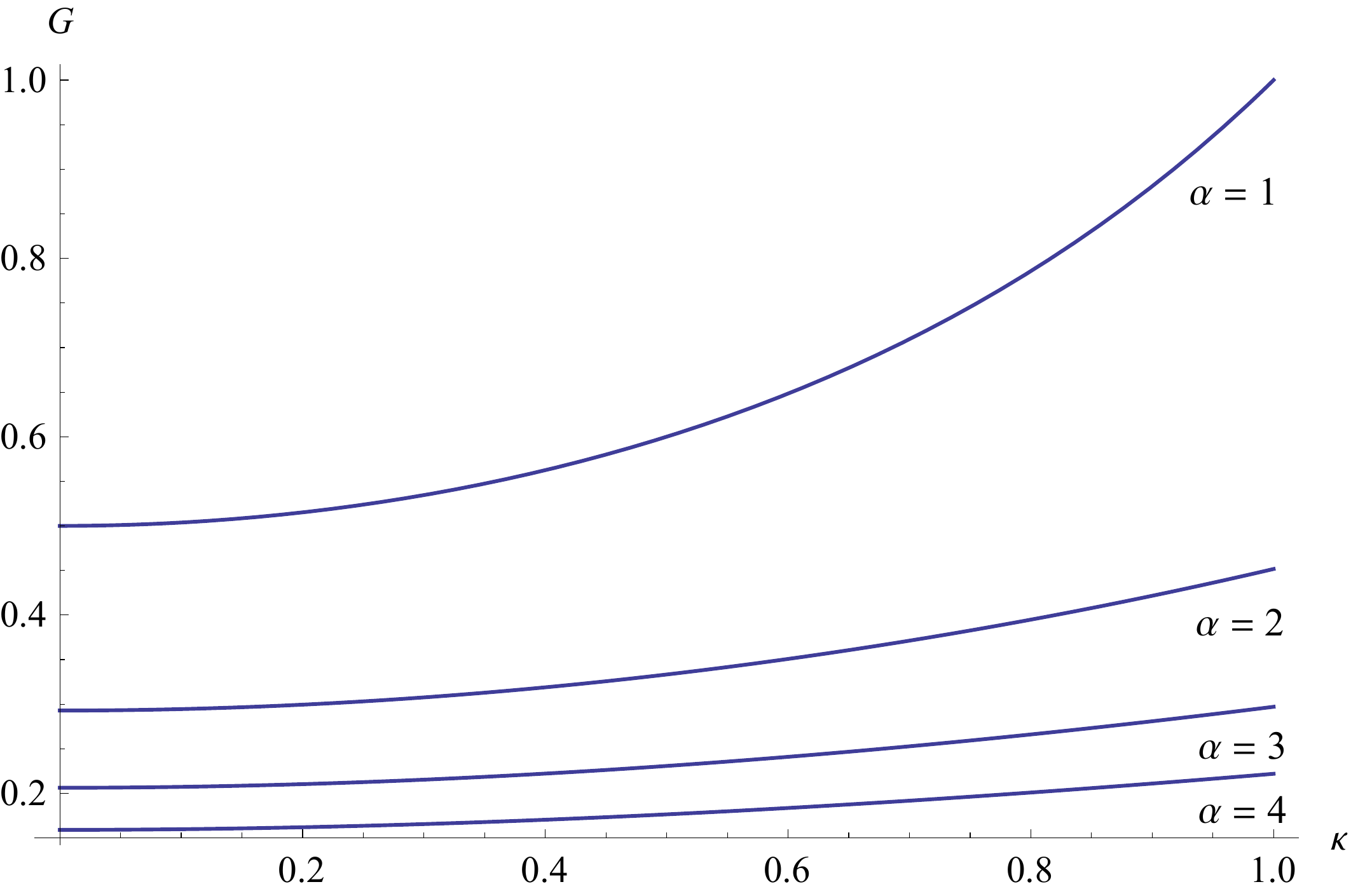}
\caption{Dependence of the Gini index of the Kaniadakis distribution on the parameters $\alpha$ and $\kappa$,
for $\kappa$ in the range $[0,1]$ and the four values $\alpha=1,2,3,4$.}
\label{Gini-Kaniadakis-4lines}
\end{center}
\end{figure}

Finally, it is interesting to compare the economic mobility defined and computed in the previous sections 
with the corresponding intuitive notion applied to a gas. We can base this comparison on the concept of temperature. 
Independently from its exact definition, mobility should be related to the probability that molecules pass to a state of higher energy 
after interaction with other molecules. It seems therefore reasonable to assume that mobility increases with absolute temperature.
In order to establish a correlation with inequality, we should then determine how the ``Gini index of a gas'' varies with temperature. 
For the Boltzmann-Gibbs exponential distribution, the Gini index is independent from the temperature and equal to $0.5$. 
A particular attention is deserved by the $\kappa$-generalized distribution introduced by Kaniadakis
\cite{Kaniadakis}. 
It takes into account relativistic kinematics
and is therefore suitable to describe hot gases and plasmas,
and
at the same time it
provides an excellent fit for income distributions both of real world
\cite {ClementiDiMGalKan}
and of kinetic models \cite{Bertotti M.L. Modanese G. 2}.
The Kanidakis distribution has a Gini index $G$ that increases with temperature.
The analytical formula which gives $G$ as a function of the parameters $\alpha$ and $\kappa$ of the distribution
contains four Gamma-functions and is given in \cite{ClementiGalKan1}. When $\alpha=1$ and $\kappa=0$,
the distribution reduces to an exponential function with $G=0.5$. For any other value of $\alpha$, $G$ is an increasing function
of $\kappa$ (compare the examples of Fig. $\ref{Gini-Kaniadakis-4lines}$). When the Kaniadakis distribution is applied to a relativistic gas,
$\kappa$ is related to the gas parameters by
\begin{equation}
\frac{1}{\kappa^2} = 1 + \frac{mc^2}{k_B T},
\label{kappa-T}
\end{equation}
where $m$ is the atomic or molecular mass, $c$ the speed of light, $k_B$ the Boltzmann constant and $T$ the absolute temperature \cite{Kaniadakis}.
Eq.\ (\ref{kappa-T}) shows that $\kappa$ increases when $T$ increases. It follows that a higher temperature leads to higher $G$.
This amounts, for the gas, to a negative correlation between equality and mobility, which disagrees with the results of empirical economic studies
and of our kinetic model.  
 
This comparison points to one of the many qualitative differences between physical and economical interactions. 
In physical systems it happens quite often that larger gradients in density, pressure, etc. cause a quicker dynamical evolution, so that, in some sense, 
``greater inequality causes greater mobility''. In economy this ``stimulus effect'' has been theorized to work for certain growth phases,
see \cite{Aghion} 
and references therein,
but is apparently not the dominant factor at equilibrium. 
Instead, mobility is positively related to equality, at least as an average over all income classes. 
We have seen that the mobility of the low and middle classes is the key element under this respect.

\section{Conclusions}
In this paper microscopic models of economic exchanges, taxation and redistribution, described by kinetic equations are considered.
Some features of the emerging income distribution curves are discussed. In particular, we compare
models with fiscal systems characterized by a different spread between the minimum and the maximum tax rate
and models in which the welfare provision is weighted differently  for different income classes.
The main contribution of the paper is represented, in our opinion, by the introduction of a measure of social mobility.
This is possible due to the structure of the equations, which involve several transition probabilities for the passage from an income class to others.
Together with the Gini index, this measure serves as a tool
to investigate the correlation between economic inequality and social mobility.
The correlation turns out to be clearly negative: a lower mobility corresponds to a higher inequality. 
This is in agreement with what is observed in real world situations. While this negative correlation has been 
the object of several studies and debates in the socio-economic arena, to our knowledge
it has not been treated within the brand of mathematical and physical approaches to complex systems.

\bigskip
\bigskip
\bigskip



\normalsize

\end{document}